\documentclass[11pt,twoside]{article}

\usepackage{asp2006}
\usepackage{epsf}
\usepackage{psfig}
\usepackage{lscape}

\begin{document}
\title{Lyman Break Galaxies and Luminous IR Galaxies at $z \sim 1$}   
\author{D. Burgarella$^1$, E. Le Floc'h$^2$, T. T. Takeuchi$^3$, V. Buat$^1$, J.S. Huang$^4$, G.H. Rieke$^2$ and K.D. Tyler$^2$}   
\affil{$^1$ Observatoire Astronomique Marseille Provence/LAM, France; $^2$ Steward Observatory, University of Arizona, USA, $^3$ Tohuku University, Japan, $^4$ Harvard-Smithsonian Center for Astrophysics, USA}    

\begin{abstract} 
We use deep GALEX images in UV to define the first large sample of 420 Lyman Break Galaxies at $z \sim 1$.  Deep Spitzer IRAC and MIPS provide the first detection of a large sample of Lyman Break Galaxies in the mid- to far-infrared range. We are therefore able to study and compare the UV and TIR emission of Lyman Break Galaxies. We find that about 15 \% of the LBG sample are strong emitters at $24 \mu m$ (Red LBGs). Most of them are Luminous IR Galaxies (LIRGs) while the rest are undetected at the $83 \mu Jy$ level of MIPS GTO image. The latter objects have a Spectral Energy Distribution similar to high redshift Lyman Break Galaxies.
\end{abstract}


\section{Introduction}   

Lyman Break Galaxies (LBGs) at high redshift are selected via a color-color plane (Steidel \& Hamilton 1993) by assuming a star formation history and a maximum amount of ultraviolet (UV) dust attenuation of about $A_{1600}=1.7$ (Giavalisco 2002). However, even though we know that some dust is present in LBGs, a direct estimate of the amount of dust is difficult because only a couple of LBGs at $z > 2$ are detected in the far infrared (FIR) and in the sub millimeter (submm) ranges (e.g. Chapman et al. 2000). Even $SPITZER$ did not detect but a handful $z > 2$ LBGs in the regime where the dust emission is the main component of the spectrum (e.g. Huang et al. 2005). Adelberger \& Steidel (2000) concluded that the bulk of the submm background is produced by moderately obscured galaxies similar to the ones already detected in UV-selected surveys. Most of their Total InfraRed (TIR) to UV luminosity ratios ($L_{TIR} / L_{UV}$) are estimated from the observed optical by using the UV slope vs. $L_{TIR} / L_{UV}$ relation defined by Meurer, Heckman \& Calzetti (1999). However, the validity of this law is questioned for normal UV-selected and FIR-selected galaxies (Bell 2002, Buat et al. 2005). It does not seem to be applicable to FIR-bright galaxies such as luminous IR galaxies (LIRGs; Goldader et al. 2002). Our approach, here, is to observe LBGs at lower redshift. The present sample was defined by Burgarella et al. (2006) in the redshift range $0.9 \leq z \leq 1.3$ from $GALEX$ in the UV and part of them observed with $Spitzer/MIPS$ in the IR. Being a true Lyman Break sample at $z \sim 1$, the selection is not exactly the same than LBGs at $z \ge 3$ because we do not assume a maximum dust attenuation and because the effect of the intergalactic medium is much lower at $z \sim 1$ than at higher redshift. However, even though differences exist in the selection process, the main objectives at all redshifts is to define a UV-selected sample with high star formation rates. We assume a cosmology with $H_0 = 70 ~km.s^{-1}.Mpc^{-1}$, $\Omega_M = 0.3$ and $\Omega_{VAC}=0.7$ in this paper.

\section{Definition of the LBG Sample}

We used DAOPHOT to make a full use of the deeper image in NUV available in GALEX Release 2 (GR2). We re-analyse (detection and flux measurement) the GR2 image of the CDFS. We restrict ourselves to the LBG sample in the redshift range $0.9 \le z \le 1.3$. Using DAOPHOT ADDSTAR task, we estimated the 80\% completeness limit to NUV=26.2 which corresponds to 45366 objects in GALEX field of view. Next, we measured the FUV flux of those NUV-selected objects and we performed a cross-correlation with COMBO 17 (Wolf et al. 2004) within a radius r=2 arcsecs. An object with an observed color $FUV-NUV \ge 2.0$ (same as Steidel \& Hamilton 1993) is classified as a LBG if the redshift of the counterparts from COMBO 17 are in the range $0.9 \le z \le 1.3$. At this point, we have 420 LBGs (2 with 4 COMBO17 counterparts, 40 with 2 counterparts and 378 with 1 counterpart). The total density of LBGs down to $NUV=26.2$ is $\sim 3500 deg^{-2}$. Then, we carry on our cross-correlation process to complete the wavelength coverture with the ESO Imaging Survey (EIS), Spitzer/IRAC and finally with SpitzerMIPS at 24 $\mu m$ and $70 \mu m$. An interesting point is that 62 of the 420 LBGs (i.e. 15 \%) have  counterpart down to MIPS/GTO limiting flux density of 83 $\mu Jy$ and we will call them Red LBGs (RLBGs hereafter). These objects are LIRGs. Among them, only 2 are detected at $70 \mu m$ but none of them are detected at $160 \mu m$ The rest, undetected in MIPS image are called Blue LBGs (BLBGs). 

\section{The Spectral Energy Distribution of LBGs at $z \sim 1$}

We analyse the Spectral Energy Distributions (SEDs) of our two classes of LBGs. In a previous paper (Burgarella et al. 2006), we have stacked the 24 $\mu$m images of BLBGs in the first LBG sample. We estimated an average flux density of $13 \mu Jy$ at the observed wavelength of 24 $\mu$m for those BLBGs. The absolute B magnitudes $M_B$ (from the observed I band) of BLBGs and RLBGs are $<M_B (BLBG)>=-21.0 \pm 1.0$  and $<M_B(RLBG)>=-22.4 \pm 1.0$ and $<M_{1800} (BLBG)>=-20.1 \pm 1.0$  and $<M_{1800} (RLBG)>=-20.7 \pm 1.1$. To give an order of magnitude, our LBGs are more luminous than Blue Compact Galaxies (e.g. $M_B<-18.5$ from Noeske et al. 2006), BLBGs are fainter than LBGs at $z \sim 3$ (e.g. $<M_{1700} (BLBG)>=-21.0 \pm 1.0$ and RLBGs are in the same range as high redshift LBGs. Our two SEDs for BLBGs and RLBGs are compared in Fig. 1 to an average SED estimated from LBGs at $z \sim 3$ in Forster-Schreiber et al. (2004) normalised at 2750 $\AA$.The SED of RLBGs is too red in the UV and present an excess in the red part of the SED. A higher dust attenuation and/or older stellar populations are very likely at the origin of this difference. The SED of BLBGs present strong similarities with high redshift LBGs within the standard deviations. The present data suggest that LBGs at $z \sim 3$ seem to have counterparts in the $z \sim 1$ universe that would be BLBGs. Indeed, the LBG selection favours starbursting galaxies with a low dust attenuation, that is BLBGs as defined in our sample.

\begin{figure*}
\epsfxsize=13truecm\epsfbox{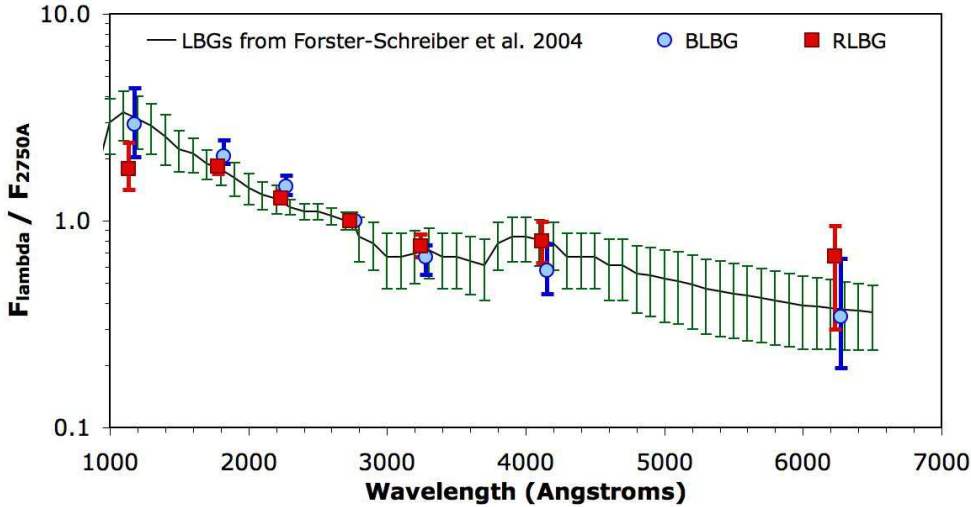}
\caption{\label{fig1} The UV/optical SEDs of the two classes of LBGs are noted as in Fig. 1 are compared to the average SED of LBGs at $z \sim 3$ by Forster-Schreiber et al. (2004). }
\end{figure*}

\section{The TIR-to-UV luminosity ratio of LBGs  at $z \approx 1$}

From the $24 \mu m$ data, we estimate the TIR-to-FUV luminosity ratio presented in Fig. 2. LBGs at $z \sim 1$ seems to have a ratio higher than Reddy et al. (2005) objects at $z \sim 2$ but lower than Buat et al. (2006) ratio at $z \sim 0$. This diagram poses the question of a possible evolution of the TIR-to-FUV luminosity ratio with the redshift.

\begin{figure*}
\epsfxsize=13truecm\epsfbox{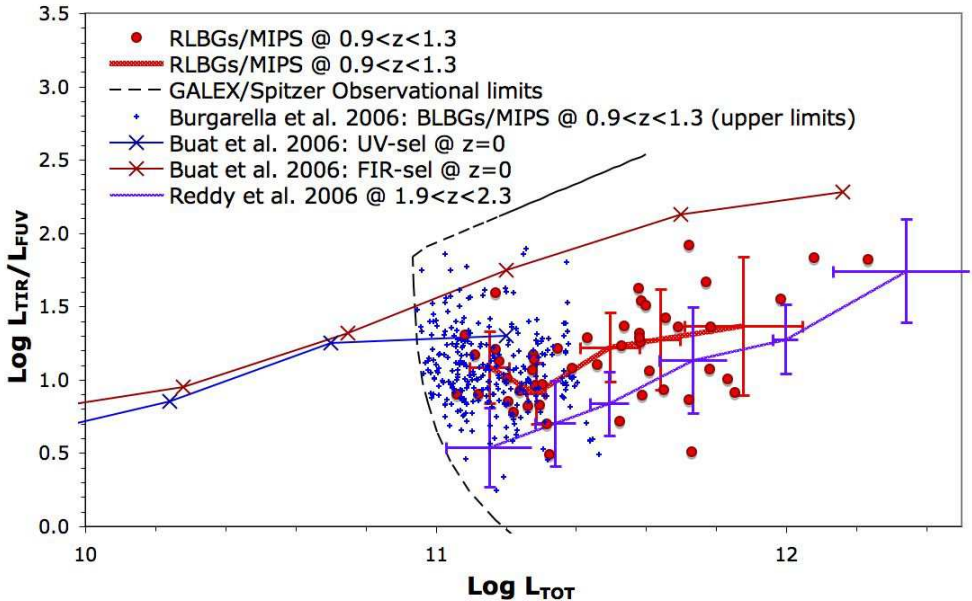}
\caption{\label{fig2} This diagram suggest that we might observe a decrease of IR/UV ratio with the redshift from z=0 to z=2.}
\end{figure*}

\section{Conclusion and Perspectives}

We analysed the spectral energy distribution of our sample of $z sim 1$ LBGs detected at $\sim 1$ by $GALEX$ and $SPITZER$. We can divide it into two sub-classes depending on whether they are detected at 24 $\mu m$: RLBGs are detected in the FIR whereas BLBGs are not. About 15 \% of our LBG sample are detected at 24 $\mu m$. Since early works carried out in the rest-frame UV on LBGs and in the rest-frame FIR on submm galaxies, the question of the Cosmic star formation history has been mainly addressed from two exclusive stand points: rest-frame UV and rest-frame IR, both of them biased by their own observational limits. Do we see the same objects in the two wavelength ranges or do they belong to two different classes well separated in space and/or in evolution ? With the detection of LBGs at $z \approx 1$ by Spitzer/MIPS, we made a first step into the multi-wavelength high redshift universe and it provides us with a clearer view of the star formation since we can simply add the UV star formation rate to the IR one to obtain an estimate of the total star formation. A full report of this work including a detailed analysis of the LBG sample (SEDs, UV luminosity functions and dust attenuation is submitted to MNRAS (Burgarella et al. 2008).

\acknowledgements 
 We  thank the French Programme National Galaxies and Programme National de Cosmologie for financial support. TTT has been supported by the 21st Century COE Program "Exploring New Science by Bridging Particle-Matter Hierarchy", Tohoku University.


\end{document}